\documentclass[prl,twocolumn,superscriptaddress,amsmath,amssymb,showpacs]{revtex4}

\usepackage{graphicx, color}

\newcommand{\ket}[1]{\ensuremath{\vert #1 \rangle}}
\newcommand{\bra}[1]{\ensuremath{\langle #1 \vert}}
\newcommand{\bracket}[2]{\ensuremath{\langle #1 \vert #2 \rangle}}
\newcommand{\ketbra}[2]{\ensuremath{\ket{#1}\bra{#1}}}

\newcommand{\ud}{\mathrm{d}}

\newcommand{\dx}{\Delta x}

\begin{document}

\title{High-fidelity fast quantum transport with imperfect controls}

\author{M.~Murphy}
\email{michael.murphy@uni-ulm.de}
\affiliation{Institut f\"{u}r Quanteninformationsverarbeitung, Universit\"{a}t Ulm, Albert-Einstein-Allee 11, 89081 Ulm, Germany}

\author{L.~Jiang}
\affiliation{Department of Physics, Harvard University, Cambridge, Massachusetts 02138}

\author{N.~Khaneja}
\affiliation{School of Engineering and Applied Sciences, Harvard University, Cambridge, Massachusetts 02138}

\author{T.~Calarco}
\affiliation{Institut f\"{u}r Quanteninformationsverarbeitung, Universit\"{a}t Ulm, Albert-Einstein-Allee 11, 89081 Ulm, Germany}

\begin{abstract}

  Effective transport of quantum information is an essential element
  of quantum computation. We consider the problem of transporting a
  quantum state by using a moving potential well, while maintaining
  the encoded quantum information. In particular, we look at a set of
  cases where the input control defining the position of the potential
  well is subject to different types of distortion, each of which is
  motivated by experimental considerations. We show that even under
  these conditions, we are able to perfectly transfer the quantum
  information non-adiabatically over any given distance.


\end{abstract}


\maketitle

In many current proposed implementations of quantum computers, it is
necessary that we can transport computational states between
operational sites. This state will often contain information from a
previous operation that we want to preserve and use in a further
operation. Consequently, we desire that the final quantum state
becomes displaced but that the initial and final states are equivalent
up to a global phase. In this sense, we will have preserved the
encoded information. In practice, this transport process is difficult
to realise without altering the state: many external sources serve to
heat or otherwise decrease the purity of the state while it is
transported. In addition, as has been pointed out in \cite{huber},
transport processes may account for 95\% of the operation time of a
quantum computation. It is therefore advantageous to minimise the time
required for this process while preserving the motional state. In this
paper, we present an analytic solution for a one-dimensional system
for transporting a quantum state over an arbitrary distance in the
non-adiabatic regime using a harmonic potential, subject to a
distortion of the input controls. This has application in systems
where the control of the transport mechanism is imperfect, as is the
case in many realistic experimental implementations, where a well
designed input control becomes distorted, either through a limitation
due to the experimental hardware, or through the interaction of the
apparatus with an uncontrolled environment. Analytic solutions to a
driven quantum harmonic oscillator have been known for many years
\cite{khandekar, husimi}, but there have been few attempts to
utilise the results for the benefit of quantum information transfer
\cite{reichle}. Our approach differs, since it is not focussed on any
one implementation for transporting quantum states (although it has
particular application to trapped particles), and we do not assume
perfect control of the system. There are also some early attempts at
high fidelity transport in experiments \cite{huber, rowe, kuhr,
  couvert}.

We first analytically solve the Schr\"odinger equation for a quantum
state confined in a driven harmonic oscillator, and show that a
suitable choice of the transport function $d(t)$ results in an
evolution of the wavepacket that constitutes what we shall refer to as
`perfect transport': the final evolved quantum state is equivalent to
the initial state up to an irrelevant global phase which is analogous
to a free evolution of the state in the frame of the potential. This
motivates us to treat the problem classically in order to derive a
particular form of the driving function that satisfies this criterion.
%
We examine the distortion of the ideal transport path analytically by
the introduction of a general functional, and show that a large class
of functionals that describe this broadening have no effect on the
success of our transport.

We model the transport of the particle from its initial position over
some arbitrary distance by a movement of the potential well according
to the functional $\mathcal{D}[d(t)]$ of an input function $d(t)$ (our
intended transport path), which defines the position of the centre of
the potential well along the axis of transport for a given time
$t$. The system can be modelled as a wave packet confined in a static
harmonic potential of fixed frequency, subject to the Hamiltonian
$\hat{H}(t) = \tfrac{1}{2}\hat{p}^2 +
\tfrac{1}{2}(\hat{x}-\mathcal{D}[d(t)])^2$, with $\hat{x}$ and
$\hat{p}$ being the usual quantum operators corresponding to position
and momentum respectively, and where $\mathcal{D}[d(t)]$ now plays the
role of a driving function. Note that we have transformed the
variables to make them dimensionless. We prepare the system in a given
eigenstate of the harmonic oscillator $\ket{\psi_n(x,t=0)}$.  If we
denote the transport distance by $\dx$, our transport condition for
the quantum case is that the fidelity between initial and final states
be $\mathcal{F} \equiv \vert
\bracket{\psi_n(x-\dx,0)}{\psi_n(x,T)}\vert^{2} = 1$; in other
words, the state be unchanged in the reference frame moving with the
potential well except for a global kinematic phase. To verify this, we
must solve the time dependent Schr\"{o}dinger equation for our
system. The normalised solutions are \cite{khandekar}
\begin{equation}\label{eqn:wf}
  \ket{\psi_n(x,t)} = e^{-i (E_n t + \tfrac{1}{2}\phi(t)-F^\prime x)}\hat{T}_F\ket{\psi_n(x,0)}\:,
\end{equation}
where $E_n$ is the energy eigenvalue corresponding to the $n$th
eigenstate, and $\hat{T}_F$ is the translation operator $\hat{T}_F
\psi(x) = \psi(x - F(t))$. The function $F(t)$ is defined as
\begin{equation}\label{eqn:f}
  F(t) \equiv \int_0^t \mathcal{D}[d(\tau)]\sin(t - \tau)\ud \tau\ ,
\end{equation}
and the phase $\phi(t)$ is given by
\begin{equation}
  \phi(t) = \int_0^t 2F(\tau)F^{\prime\prime}(\tau) + F^{\prime\prime 2}(\tau) + F^{\prime 2}(\tau)\; \mathrm{d}\tau\:.
\end{equation}
We can satisfy our transport condition if the function $F(t)$
satisfies $F(T) = \dx$, $F(0) = F^\prime(0) = F^\prime(T) = 0$, such
that when $t = 0$, the wavefunction in equation~(\ref{eqn:wf}) reduces
to the normalised eigenstates of the harmonic oscillator, and at $t =
T$, the wavefunction reduces to the original wavefunctions shifted by
an amount $\dx$ and with a global phase
$\exp[-i(E_nT+\tfrac{1}{2}\phi(T)]$.

There is a direct correspondence between the wavefunction in
equation~(\ref{eqn:wf}) and the equations of motion for the classical
analogue of our system, which can be seen by taking the expectation
values of the operators $\hat{x}$ and $\hat{p}$ of the state in
equation~(\ref{eqn:wf}), and then noting that the time evolution of
these quantities obey the Newtonian equations of motion for a
classical particle confined within a harmonic potential well with
constant frequency. The equations of motion are given by $\dot{x}(t) =
p(t)$, $\dot{p}(t) = \mathcal{D}[d(t)] - x(t)$, where again all
variables are rescaled to make them dimensionless. Here, $x(t)$ and
$p(t)$ refer respectively to the position and momentum of the
classical particle along the axis of transport at time $t$. These
equations have solutions $x(t) = x_c(t) + F(t)$ and $p(t) = p_c(t) +
F^\prime(t)$, with $F(t)$ as defined in equation~\eqref{eqn:f}. The
functions $x_c(t)$ and $p_c(t)$ are the solutions to the homogeneous
equations of motion, which therefore describe the simple harmonic
motion undergone by the particle when no transport is undertaken. In
the classical picture, the condition for performing perfect transport
becomes $x(T) = x_c(T) + \dx$ and $p(T) = p_c(T)$, where $\dx$ denotes
the displacement of the potential well at the final time. Our perfect
transport condition specifies the boundary conditions on $F(t)$ and
its first-order derivatives. We can rewrite $F(t)$ in terms of
$\mathcal{D}[d(t)]$ by noting that equation~\eqref{eqn:f} is a
Volterra integral equation of the first kind with a trigonometric
kernel \cite{hie}, and hence has the solution
\begin{equation}\label{eqn:ddef}
  \mathcal{D}[d(t)] = F^{\prime\prime}(t)+F(t)\ .
\end{equation}
We first search for a form for the function $d(t)$ by
supposing that $\mathcal{D}[d(t)] = d(t)$. Substituting the transport
conditions on $x(t)$ and $p(t)$ into equation~\eqref{eqn:ddef} and
imposing the condition $F^{\prime\prime}(0) = F^{\prime\prime}(T) = 0$,
we find $F(0) = 0$ and $F(T) = \dx$. If we fix the condition
that $d^\prime(t) = 0$ when $t = 0$ and $T$, we subsequently place
boundary conditions on $F^{\prime}(t)$ and
$F^{\prime\prime\prime}(t)$ at the initial and final times. Taking all
of these conditions into account for $F(t)$ and its derivatives, we
can now construct a function $d(t)$ that transports our particle
perfectly. This procedure is as follows: we construct a general $F(t)$
by taking the simplest form of transport function (a linear function)
and adding a series of Fourier components.  We scale the components so
that their period matches the transport time. We then apply the
boundary conditions on $F(t)$ to solve for the Fourier
coefficients. Due to the periodicity of the components, we have only
five independent boundary conditions, and so we may uniquely specify
only this many Fourier components. Substituting this into
equation~\eqref{eqn:ddef} gives us the solution
\begin{equation}
\begin{split}\label{eqn:opt}
  d_o(t) = & \dx \biggl[ \frac{t}{T} + \sin\left(\frac{2\pi t}{T}\right)\cdot
    \left(\frac{8\pi}{3T^2}-\frac{2}{3\pi}\right) + \\ & +
    \sin\left(\frac{4\pi t}{T}\right)\cdot\left(\frac{1}{12\pi}-\frac{4\pi}{3T^2}\right)\biggr]\ ,
\end{split}
\end{equation}
which we shall call our transport function.  This equation depends on
the transport time $T$, so that one must choose the correct transport
function for the appropriate transport time. Since we have scaled all
the variables in our system, $T = 2\pi$ represents transport over one
period of the harmonic motion in the potential well (the trap
period). We henceforth assume that $d(t) \equiv d_o(t)$. (It should be
noted that one can follow a more rigorous derivation by starting from
a description in the frame of an optimal control problem, but this was
not presented here. See for instance \cite{brockett}.)

Through consideration of the above, we may state the following: if the
functional $\mathcal{D}[d(t)]$ has a form such that the boundary
conditions on $F(t)$ are preserved (and $\mathcal{D}[d(t)]$ is
non-singular for all $t$), then the functional $\mathcal{D}$ does not
affect the transport of the particle. We consider the following three
forms for $\mathcal{D}[d(t)]$, and briefly discuss the motives for
doing so.

(i) The $\dot{d}(t)$ model: $\mathcal{D}[d(t)] = d(t) +
\alpha\dot{d}(t)$, where $\alpha$ is a real constant, and $\dot{d}(t)$
represents differentiation with respect to time. A physical
interpretation for this model could be that we consistently
`overshoot' or `undershoot' our desired potential well position, so
that as we move the well more quickly, the deviation from the desired
position becomes greater.

(ii) The \emph{piecewise} model: The functional $\mathcal{D}$
casts $d(t)$ into a piecewise form
\begin{equation}
\label{eqn:opt_piece}
  \mathcal{D}[d(t)] = d(t_n)\ \text{for}\
  t \in [t_n-\tfrac{T}{2N}, t_{n}+\tfrac{T}{2N}]\ ,
\end{equation}
where $t_n = \tfrac{nT}{N}$ for a given $N \in \mathbb{Z}^+,\ N >
1$. This has the effect that the potential undergoes discrete `jumps'
in its position along the transport axis, which could be due to a
sampling rate limitation of the equipment used for experiment.

(iii) The \emph{Fourier} model: $\mathcal{D}[d(t)] = d(t) + g(t)$,
where $g(t)$ is a discrete Fourier series. This is of relevance since
we may decompose any periodic signal (for instance, periodic pulse
distortions) into a Fourier series (note that the Fourier
representation must converge at all times $t$ to the signal being
represented).
\begin{figure}[tbp]
  \includegraphics{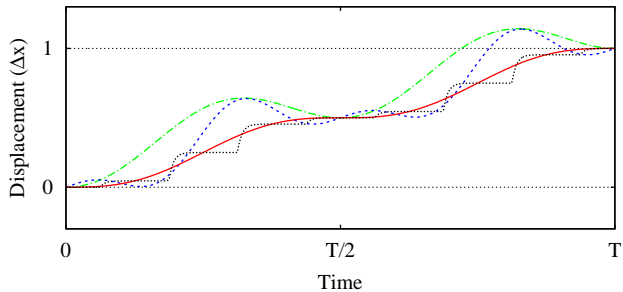}
  \caption{(Colour online) Distortion of the transport function over
    time period $T = 2\pi$.  The thick (red) solid line is the
    transport function from equation~(\ref{eqn:opt}), dot-dashed
    (green) line is the $\dot{d}(t)$ model ($\alpha = 1$), the dashed
    (blue) line is the Fourier model (random coefficients in the range
    [-1,1]), and the dotted (black) line is the piecewise model with
    an exponential-type smoothing ($N = 8$).}
  \label{fig:homogb}
\end{figure}

We now show that each of these functionals the conditions for perfect
transport are still satisfied. In the case of the $\dot{d}(t)$ model, we begin
by writing $d(t) = F^{\prime\prime}(t) + F(t)$. Hence
$\mathcal{D}[d(t)] = \tilde{F}^{\prime\prime}(t) + \tilde{F}(t)$,
where $\tilde{F}(t) = F^\prime(t) + F(t)$. It should be clear that
$\tilde{F}(t)$ satisfies the same boundary conditions as $F(t)$. Hence
we conclude that we achieve our desired transport with this functional
for any transport time $T$.

In the case of the piecewise function, we substitute the functional
directly into equation~\eqref{eqn:f} and solve for $F(t)$. We find in
the limit that $T \rightarrow T_k = 2 k \pi,\ k \in \mathbb{Z}^+$, the
boundary conditions for $F(t)$ are satisfied as before. In other
words, the piecewise functional can only achieves the boundary
conditions when the transport time is an integer multiple of the
period of the harmonic trap. It may additionally be the case that the
movement of the potential is not exactly stepwise, but that instead
the movement is smoothed out (for instance, if we consider a segmented
ion trap, this will be due to the charging characteristics of the
electrodes). We can model this by writing
equation~(\ref{eqn:opt_piece}) as
\begin{equation}
\begin{split}
\label{eqn:opt_piece2}
   \mathcal{D}[d(t)] = d(t_n) -
   q(t-t_n+\tfrac{T}{2N})\bigl[d(t_n) - d(t_{n-1})\bigr]\ \\
\end{split}
\end{equation}
so that $q(t)$ describes the smoothing from the previous value in the
stepwise function to the next. Substituting this into
equation~(\ref{eqn:f}), we can calculate that the part of the integral
dependent on $q(t)$ evaluates to zero. Hence we may conclude that any
smoothing of the transport path due to these terms may be ignored.

Finally, we consider the Fourier model. Again, through direct
integration of the functional via equation~\eqref{eqn:f}, we can
obtain the associated function $F(t)$. Here, we see that we satisfy
the boundary conditions if the period of the function $g(t)$ is $T_k /
2$, with $T_k$ as given above. We can satisfy this by tuning the the
frequency of the harmonic potential to accommodate the periodic noise.
\begin{figure}[tbp]
  \includegraphics{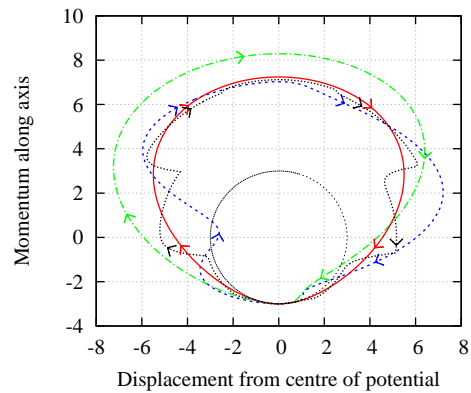}
  \caption{(Colour online) Trajectories through phase space in the
    frame of the moving potential of a transported particle over time
    period $T = 2\pi$. See caption of Figure~\ref{fig:homogb} for
    key. For illustative purposes, the particle was given some initial
    momentum. The centre circle is the trajectory of a freely
    oscillating particle with constant energy and the same momentum in
    a static potential.}
  \label{fig:phasespace}
\end{figure}
Figure~\ref{fig:homogb} shows sample transport paths $\mathcal{D}[d(t)]$
over a short transport time $T = 2\pi$. One can see that the deviation
here is not small; we significantly disturb the motion of the
particle. Figure~\ref{fig:phasespace} shows the classical trajectories
through phase space of the particles transported according to the
transport paths $\mathcal{D}[d(t)]$ (or, equivalently, the expectation
values of the operators $\hat{x}$ and $\hat{p}$). One sees here that
the trajectories begin and end on the constant energy curve given by
the free oscillation in the well.

\begin{figure}[tbp]
  \includegraphics{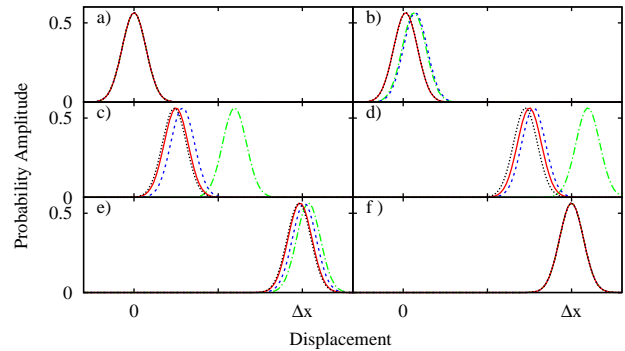}
  \caption{(Colour online) The evolution of the ground state
    probability distribution from (a) an initial time $t=0$ to (f) the
    final time $T = 2\pi$. See caption of Figure~\ref{fig:homogb} for
    key.}
  \label{fig:quantra}
\end{figure}

Figure~\ref{fig:quantra} shows snapshots of the evolution of
the ground state wavefunction subject to distortion;
although the probability distributions diverge from each other at
intermediate times for the different models for $\mathcal{D}[d(t)]$,
at the final time the wavefunctions converge to the initial profile of
the ground state probability distribution displaces to the final
position. Furthermore, one may verify that the energy expectation at
the final time is the predicted value of $\langle E \rangle = E_n = (n
+ \tfrac{1}{2})$, which can be immediately found from
equation~(\ref{eqn:wf}) (although $\langle E \rangle$ is only uniquely
specified for the ground state).

\begin{figure}[tbp]
  \includegraphics{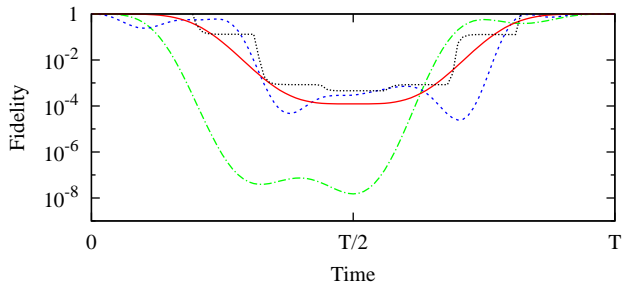}
  \caption{(Colour online) The fidelity between the instantaneous
    ground state and the transported ground state over a time $T =
    2\pi$. See caption of Figure~\ref{fig:homogb} for key. }
  \label{fig:fidel_adiab}
\end{figure}

The scheme also allows us to transport far from the adiabatic limit,
which can be demonstrated by calculating the fidelity between the
actual transported ground state and the instantaneous ground state of
the displaced potential well for different values of the transport
time. As we transport the state, the fidelity deviates further from
unity at intermediate times, only to recover again at the final time
(demonstrated in Figure~\ref{fig:fidel_adiab}).
Although here we have chosen the ground state of the potential, we may
well have used any of the other eigenstates of the harmonic oscillator
and produced similar results.

We can also begin with superpositions of states. Consider the coherent
superposition of eigenstates given by $\ket{\Psi(x,0)} = \sum_{n=0}^M
c_n \ket{\psi_n(x,0)}$, where $c_n$ are normalised
coefficients. In order to recover the maximum fidelity, we must
preserve the relative phases between initial states. The $n$th
eigenstate acquires a relative phase $\exp[-i(n-m)T]$ with the $m$th
eigenstate during transport, and so in order to preserve its phase
relation with the other superposed states, we can choose $T =
T_k$. Hence the transport condition is satisfied.

If we instead begin with a mixed state described by the density matrix
$\rho(x,0) = \sum_{n=0}^M \rho_n \ketbra{\psi_n}{\psi_n}$, and take
the fidelity\cite{jozsa} $\mathcal{F} =
(\mathrm{tr}[\sqrt{\rho_1}\rho_2\sqrt{\rho_1}]^{1/2})^2$ where
$\rho_1$ and $\rho_2$ are the initial and final states respectively,
we can show that the fidelity of transport does not depend on the
coefficients $\rho_n$, so that the distribution of states remains
constant during transport. We can then infer that the transport is
insensitive to temperature.

In this paper, we have derived analytic solutions for the transport of
a quantum state via a moving harmonic potential with a constant
frequency. We considered the conditions under which a functional that
distorts the input control achieves the conditions we have set for
perfect transport. In particular, we studied three different models
that have a quantitative relevance in experiments dealing with quantum
transport, and showed that under certain conditions, all three models
describe a broadening of the transport path without detriment to the
transport success. We briefly review these conditions. (i) The
$\dot{d}(t)$ model fulfills the transport condition for any transport
time. (ii) The piecewise model fulfills the transport condition when
the transport time is an integer multiple of the trap period. (iii)
The Fourier model (which models signal distortions) satisfies the
boundary conditions if the signal is periodic with half the trap
period.

Of course, in a realistic situation beyond our harmonic oscillator
model, complete insensitivity to such a broad range of control
imperfections is not to be expected: in that case, we not only have
distortion of the input controls, but also distortion of the shape of
the potential itself. However, our result indicates significant
robustness (by which we mean a low sensitivity of the transport fidelity
to such distortions) is likely to be obtained at least when potential
anharmonicities are small, since the system is well approximated by
the harmonic oscillator. This will be the subject of future
investigations. Such deviations may also be overcome by application of
optimal control methods \cite{calarco, tannor2}. The consequence of
performing robust transport, particularly in the presence of such
imperfections, is that we can distribute quantum information to
separate elements of a quantum computer with a very high fidelity in
short times and over large distances. This is essential for
scalability and fault tolerance of quantum systems for future use in
computation.

We acknowledge support by the European Union under the contract
MRTN-CT-2006-035369 (EMALI) and the Integrated Project SCALA.

\bibliography{paper_final}

\begin{thebibliography}{12}
\expandafter\ifx\csname natexlab\endcsname\relax\def\natexlab#1{#1}\fi
\expandafter\ifx\csname bibnamefont\endcsname\relax
  \def\bibnamefont#1{#1}\fi
\expandafter\ifx\csname bibfnamefont\endcsname\relax
  \def\bibfnamefont#1{#1}\fi
\expandafter\ifx\csname citenamefont\endcsname\relax
  \def\citenamefont#1{#1}\fi
\expandafter\ifx\csname url\endcsname\relax
  \def\url#1{\texttt{#1}}\fi
\expandafter\ifx\csname urlprefix\endcsname\relax\def\urlprefix{URL }\fi
\providecommand{\bibinfo}[2]{#2}
\providecommand{\eprint}[2][]{\url{#2}}

\bibitem[{\citenamefont{Huber et~al.}(2008)\citenamefont{Huber, Deuschle,
  Schnitzler, Reichle, Singer, and Schmidt-Kaler}}]{huber}
\bibinfo{author}{\bibfnamefont{G.}~\bibnamefont{Huber}},
  \bibinfo{author}{\bibfnamefont{T.}~\bibnamefont{Deuschle}},
  \bibinfo{author}{\bibfnamefont{W.}~\bibnamefont{Schnitzler}},
  \bibinfo{author}{\bibfnamefont{R.}~\bibnamefont{Reichle}},
  \bibinfo{author}{\bibfnamefont{K.}~\bibnamefont{Singer}}, \bibnamefont{and}
  \bibinfo{author}{\bibfnamefont{F.}~\bibnamefont{Schmidt-Kaler}},
  \bibinfo{journal}{New Journal of Physics} \textbf{\bibinfo{volume}{10}},
  \bibinfo{pages}{013004} (\bibinfo{year}{2008}).

\bibitem[{\citenamefont{Khandekar and Lawande}(1979)}]{khandekar}
\bibinfo{author}{\bibfnamefont{D.~C.} \bibnamefont{Khandekar}}
  \bibnamefont{and} \bibinfo{author}{\bibfnamefont{S.~V.}
  \bibnamefont{Lawande}}, \bibinfo{journal}{Journal of Mathematical Physics}
  \textbf{\bibinfo{volume}{20}}, \bibinfo{pages}{1870} (\bibinfo{year}{1979}).

\bibitem[{\citenamefont{Husimi}(1953)}]{husimi}
\bibinfo{author}{\bibfnamefont{K.}~\bibnamefont{Husimi}},
  \bibinfo{journal}{Prog. Theo. Phys.} \textbf{\bibinfo{volume}{9}},
  \bibinfo{pages}{381} (\bibinfo{year}{1953}).

\bibitem[{\citenamefont{{R. Reichle \emph{et al.}}}(2006)}]{reichle}
\bibinfo{author}{\bibnamefont{{R. Reichle \emph{et al.}}}},
  \bibinfo{journal}{Fortschritte der Physik} \textbf{\bibinfo{volume}{54}},
  \bibinfo{pages}{666} (\bibinfo{year}{2006}).

\bibitem[{\citenamefont{{M.A. Rowe \emph{et al.}}}(2002)}]{rowe}
\bibinfo{author}{\bibnamefont{{M.A. Rowe \emph{et al.}}}},
  \bibinfo{journal}{Quantum Information and Computation}
  \textbf{\bibinfo{volume}{2}}, \bibinfo{pages}{257} (\bibinfo{year}{2002}).

\bibitem[{\citenamefont{{S. Kuhr \emph{et al.}}}(2003)}]{kuhr}
\bibinfo{author}{\bibnamefont{{S. Kuhr \emph{et al.}}}},
  \bibinfo{journal}{Phys. Rev. Lett.} \textbf{\bibinfo{volume}{91}},
  \bibinfo{pages}{213002} (\bibinfo{year}{2003}).

\bibitem[{\citenamefont{Couvert et~al.}(2008)\citenamefont{Couvert, Kawalec,
  Reinaudi, and Gu\'{e}ry-Odelin}}]{couvert}
\bibinfo{author}{\bibfnamefont{A.}~\bibnamefont{Couvert}},
  \bibinfo{author}{\bibfnamefont{T.}~\bibnamefont{Kawalec}},
  \bibinfo{author}{\bibfnamefont{G.}~\bibnamefont{Reinaudi}}, \bibnamefont{and}
  \bibinfo{author}{\bibfnamefont{D.}~\bibnamefont{Gu\'{e}ry-Odelin}},
  \bibinfo{journal}{Europhys. Lett.} \textbf{\bibinfo{volume}{83}},
  \bibinfo{pages}{13001 (4pp)} (\bibinfo{year}{2008}).

\bibitem[{\citenamefont{Polyanain and Manzhirov}(1998)}]{hie}
\bibinfo{author}{\bibfnamefont{A.~D.} \bibnamefont{Polyanain}}
  \bibnamefont{and} \bibinfo{author}{\bibfnamefont{A.~V.}
  \bibnamefont{Manzhirov}}, \emph{\bibinfo{title}{Handbook of Integral
  Equations}} (\bibinfo{publisher}{CRC Press, Boca Raton},
  \bibinfo{year}{1998}).

\bibitem[{\citenamefont{Brockett}(1970)}]{brockett}
\bibinfo{author}{\bibfnamefont{R.~W.} \bibnamefont{Brockett}},
  \emph{\bibinfo{title}{Finite Dimensional Linear Systems}}
  (\bibinfo{publisher}{Wiley New York}, \bibinfo{year}{1970}).

\bibitem[{\citenamefont{Jozsa}(1994)}]{jozsa}
\bibinfo{author}{\bibfnamefont{R.}~\bibnamefont{Jozsa}},
  \bibinfo{journal}{Journal of Modern Optics} \textbf{\bibinfo{volume}{41}},
  \bibinfo{pages}{2315} (\bibinfo{year}{1994}).

\bibitem[{\citenamefont{Calarco et~al.}(2004)\citenamefont{Calarco, Dorner,
  Julienne, Williams, and Zoller}}]{calarco}
\bibinfo{author}{\bibfnamefont{T.}~\bibnamefont{Calarco}},
  \bibinfo{author}{\bibfnamefont{U.}~\bibnamefont{Dorner}},
  \bibinfo{author}{\bibfnamefont{P.~S.} \bibnamefont{Julienne}},
  \bibinfo{author}{\bibfnamefont{C.~J.} \bibnamefont{Williams}},
  \bibnamefont{and} \bibinfo{author}{\bibfnamefont{P.}~\bibnamefont{Zoller}},
  \bibinfo{journal}{Phys. Rev. A} \textbf{\bibinfo{volume}{70}},
  \bibinfo{eid}{012306} (\bibinfo{year}{2004}).

\bibitem[{\citenamefont{Sklarz and Tannor}(2002)}]{tannor2}
\bibinfo{author}{\bibfnamefont{S.~E.} \bibnamefont{Sklarz}} \bibnamefont{and}
  \bibinfo{author}{\bibfnamefont{D.~J.} \bibnamefont{Tannor}},
  \bibinfo{journal}{Phys. Rev. A} \textbf{\bibinfo{volume}{66}},
  \bibinfo{eid}{053619} (\bibinfo{year}{2002}).

\end{thebibliography}

\end{document}